\documentstyle[emulateapj,psfig]{article}  
\slugcomment{ApJ Letter, 2000, 531, in press}  
\lefthead{ZHANG, HARDING \& MUSLIMOV}  
\righthead{RADIO PULSAR DEATH LINE REVISITED}  
 
\begin{document}   
\title{Radio pulsar death line revisited: is PSR J2144-3933 anomalous?}   
\author{Bing Zhang\altaffilmark{1,2}, Alice K. Harding\altaffilmark{1}
 and Alexander G. Muslimov\altaffilmark{3}}   
\altaffiltext{1} {Laboratory of High Energy Astrophysics, NASA/Goddard Space  
 Flight Center, Greenbelt, MD 20771}
\altaffiltext{2} {National Research Council Research Associate Fellow}  
\altaffiltext{3} {SM\&A Corporation, Space Sciences Division, Upper
Marlboro, MD 20774}  

\begin{abstract}
We reinvestigate the radio pulsar ``death lines'' within the framework 
of two different types of polar cap acceleration models, i.e., the
vacuum gap model and the space-charge-limited flow model, with either
curvature radiation or inverse Compton scattering photons as the source 
of pairs. General relativistic frame-dragging is taken into account in
both models. We find that the inverse Compton scattering induced 
space-charge-limited flow model can sustain strong pair production 
in some long-period pulsars, which allows the newly detected 
8.5s pulsar PSR J2144-3933 to be radio loud, without assuming a special 
neutron star equation-of-state or ad hoc magnetic field configurations.
\end{abstract}   
   
   
\section{Introduction}
Copious pair production in pulsar inner magnetospheres has long been 
conjectured as an essential condition for the radio emission of 
pulsars since the pioneering work of Sturrock (1971)\footnote{Although 
some authors argued that pair production might not be an essential 
condition for pulsar radio emission (e.g. Weatherall \& Eilek 1997),
no such pairless radio emission theory is fully constructed.}. 
Generally speaking, pair production can create a dense plasma which 
may allow various coherent instabilities to grow, so as to account 
for the high brightness temperature observed from the pulsars. 
The secondary pairs (rather than primary particles) with $\gamma\sim 
10^2-10^4$ can produce typical radio-band emission within various models 
(e.g. Ruderman \& Sutherland 1975, hereafter RS75; Melrose 1978; Qiao 
\& Lin 1998).
As a consequence, the so-called radio pulsar ``death line'' is 
defined as a line in a two-dimensional pulsar parameter phase space
($P-\dot P$ diagram, $P-B_s$ diagram, or $P-\Phi$ diagram, where
$P$ is the pulsar period, $\dot P$ the pulsar spin-down rate, $B_s$ 
the surface magnetic field, and $\Phi$ the polar cap potential),
which separates the pulsars which can support pair production in 
their inner magnetospheres from those which cannot (RS75; Arons
\& Scharlemann 1979, hereafter AS79; Chen \& Ruderman 1993, hereafter 
CR93; Rudak \& Ritter 1994; Qiao \& Zhang 1996; Bj\"ornsson 1996; 
Weatherall \& Eilek 1997; Arons 1998, 2000). Present pulsar death line
theories require some degree of anomalous field line configurations 
(multipole component or offset dipole) to interpret known pulsar data.
However, a recently discovered long-period (8.5s) pulsar PSR J2144-3933 
(Young, Manchester \& Johnston 1999) is clearly located well beyond 
the conventional death valley (CR93), unless special neutron star 
equation-of-state or even ad hoc magnetic field configurations are 
assumed. This challenges the widely accepted belief that pair 
production is essential for pulsar radio emission. Arons (1999) 
found that this pulsar is located within the death valley of an
inverse Compton controlled space-charge-limited flow model with 
frame-dragging included, but some degree of the point dipole offset 
is needed.

\section{Death lines in various models}

The death line has been defined (e.g. RS75) by the condition
that the potential drop across the accelerator ($\Delta V$) 
required to produce enough pairs per primary to screen out the 
parallel electric field 
is larger than the maximum potential drop 
($\Phi_{\rm max}$) available from the pulsar, 
in which case no secondary pairs would be produced
\footnote{Note that 
there are some additional criteria to constrain the death lines for 
the millisecond pulsars (Rudak \& Ritter 1994; Qiao \& Zhang 1996; 
Bj\"ornsson 1996) and the pulsars with strong surface magnetic fields 
(Baring \& Harding 1998), and we will not address them in this paper.}.

It is worth noting that pulsar death lines are actually 
{\em model-dependent}. Besides the model-dependent $\Phi_{\rm max}$
(see [\ref{Phimax1}] and [\ref{Phimax2}]), the form of $E_\parallel$ 
within the accelerator (which depends on the boundary conditions 
and on whether the general relativistic frame-dragging
effect is taken into account), the typical energy of the 
$\gamma$-ray photons (which depends on whether their origin is 
curvature radiation (CR) or inverse Compton scattering (ICS)), and 
the strength of the perpendicular magnetic field the $\gamma$-ray 
photon encounters (which depends on the field strength and the field 
line curvature near the neutron star surface) can change $\Delta V$
considerably and alter the death lines.
Furthermore, the obliquity of the pulsar (which changes 
$E_\parallel$ and $\Phi_{\rm max}$) and the equation-of-state of the 
neutron star (which changes the moment of inertia $I=10^{45}
I_{45}$ and the radius $R=10^6 R_6$ of the star)
will also influence the location of the death lines. As a result, 
the phase space for pulsars to die should be a ``valley'' rather 
than a single line. CR93 defined a death valley, within the framework 
of the RS75 vacuum gap model, as the phase space range between the death 
line of the dipolar field configuration and the death line of some special 
multipolar field configurations.
Here we will also adopt such a death 
valley, and regard it as also including the scatter of obliguities for 
different pulsars\footnote{Strictly speaking, there could be different 
death lines for pulsars with different obliquities. The death lines 
could be quite different for the extreme cases of aligned and orthognal 
rotators.}. Modification of the equation-of-state
or of the polar cap radius will also modify the death valleys 
systematically. The surface magnetic field for the star-centered
dipolar configuration is $B_d=6.4\times 10^{19}(P \dot P)^{1/2}
I_{45}^{1/2}R_6^{-3}$ regardless of the internal field geometry 
(Shapiro \& Teukolsky 1983; Usov \& Melrose 1995). Thus the only model
dependence of the dipolar surface field is the offset of the dipole
center from the star center.

Two subgroups of inner gap models were proposed by adopting different
boundary conditions. The vacuum-like gap model (V model, 
RS75) assumes strong binding of ions at the neutron star surface 
($E_\parallel(z=0)\neq 0$), while the space-charge-limited flow model 
(SCLF model, AS79) assumes free 
emission of particles from the surface ($E_\parallel (z=0)=0$). Both 
of these models were originally proposed by assuming that CR of the 
primary particles is the main mechanism to create $\gamma$-ray seeds 
to ignite the pair production cascades, and by neglecting general 
relativistic effects. They were improved later by different authors 
to include ICS and the inertial frame-dragging effects. 
The role of frame-dragging in pulsar physics was first explored by
Muslimov \& Tsygan (1992, hereafter MT92) and updated by 
Muslimov \& Harding (1997, hereafter MH97)
and Harding \& Muslimov (1998, hereafter HM98) 
within the framework of the SCLF 
model. MT92 noted that since stellar rotation actually drags the local 
inertial frame (LIF) to rotate with an angular velocity $\Omega_{\rm 
LIF}\simeq \Omega_{*}\kappa_g (R/r)^3$ ($\Omega_{*}$ is the angular
velocity of the star), the electric field required to bring a charged 
particle into corotation is weaker than that 
in the flat spacetime, since such a field only needs to compensate 
the angular velocity difference between $\Omega_{*}$ and 
$\Omega_{\rm LIF}$ rather than the difference of $\Omega_{*}$ 
and the angular velocity at infinity (which is 0). As a result, 
near the star surface, the Goldreich-Julian density 
becomes\footnote{For an explicit expression for $\eta_R$, 
see eqs.[4], [8] and [16] of HM98.}
\begin{equation}
\eta_R \simeq -{{\bf (\Omega_{*}-\Omega_{\rm LIF})\cdot B} \over
2\pi c\alpha}\simeq -{{\bf \Omega_{*}\cdot B}\over 2 \pi c\alpha}
\left[1-\kappa_g \left({R\over r}\right)^3\right],
\label{etaR}
\end{equation}
where $\kappa_g\sim (r_g/R)(I/M R^2)\sim 0.15-0.27$ (HM98; we will 
adopt a typical value of 0.15 hereafter), $\alpha =(1-r_g/R)^{1/2}
\sim 0.78$ is the redshift factor, $r_g$ is the gravitational radius, 
and $M$ is the mass of the neutron star. 

\subsection{Vacuum gap model (V model)}

In principle, pulsar $E_\parallel$ arises from the deviation of the
local charge density ($\eta$) from the Goldreich-Julian density 
($\eta_{R}$). If the binding energy of the positive ions is large 
enough to prevent the ions from thermionic or field emission ejection, 
a vacuum-like gap (RS75) will form right above
the neutron star surface, in which the charge depletion is very large
(of the order of $\eta_{R}$ itself) so that a very strong $E_\parallel$
is built up right above the surface. This picture was questioned since 
later calculations showed that the ion binding energy is actually not
high enough (for recent reviews, see, e.g. Usov \& Melrose 1995). 
Recent observations indicate that some pulsars may favor the existence 
of such vacuum gaps (Vivekanand \& Joshi 1999; Deshpande \& Rankin 1999), 
and some ideas to solve the binding energy problem have been proposed
(e.g. Xu, Qiao \& Zhang 1999). The inclusion of ICS in such a model 
was carried out by Zhang \& Qiao (1996) 
and Zhang et al. (1997), and the corresponding death line was examined 
by Qiao \& Zhang (1996). The influence of the frame-dragging effect on 
the model has not been examined before.

The maximum potential available for the V model is just the homopolar
generator potential, which is the potential difference between the pole 
and the edge of the polar cap at the star surface and reads
\begin{equation}
\Phi_{\rm max}({\rm V})\simeq(1-\kappa_g){B_dR^3\Omega_{*}^2 \over 2c^2}
\simeq 5.6\times 10^{12}({\rm Volts})B_{d,12}P^{-2}R_6^3.
\label{Phimax1}
\end{equation}
Note frame-dragging modifies the flat spacetime result by a factor
of $\zeta=(1-\kappa_g)\sim 0.85$, which arises from $(\Omega_{*}
-\Omega_{\rm LIF})$. 
The solution of the one-dimensional Poisson's equation (the infinitesimal
gap in RS75) then gives $E_\parallel(z)=\zeta(2\Omega_{*}B/ c)(h-z)$ 
and $\Delta V=\zeta (\Omega_{*}B/ c)h^2$, which are essentially RS75's 
results with the correction factor $\zeta$.

The gap height depends on three length scales, 
i.e., $l_{\rm acc}$, the 
acceleration scale before the primary electron or positron achieves a 
high enough energy $\gamma_c$; $l_e\sim c[\dot\gamma(\gamma_c mc^2)/E_c]
^{-1}$, the mean free path of the electron/positron with 
$\gamma_c$ to emit one $\gamma$-ray quanta with energy $E_c$; and 
$l_{ph}=\chi\rho(B_{cri}/B_s)(2mc^2/E_c)$, the mean free path 
of the $\gamma$-photon before being absorbed, where $B_{cri}
=m^2c^3/\hbar e\simeq 4.4\times 10^{13}$G is the 
critical magnetic field, and $\chi\sim 0.1$ is the key 
parameter to describe $\gamma-B$ absorption coefficient (Erber 
1966). The electron mean free path $l_e$ should not exceed the
gap height $h=l_{\rm acc}+l_{ph}$. For the V model, one gets
$\gamma_c=2\zeta(e/ mc^2)(\Omega_{*} B/ c)l_{\rm acc}
(h-l_{\rm acc}/2)$ with the form of $E_\parallel$. For the CR-induced 
cascade model, we have $E_c=(3/2)(\hbar c/\rho)\gamma_c^3$, and 
$l_e=(9/4)(\hbar c/e^2)\rho\gamma_c^{-1}\ll l_{ph}\sim l_{\rm acc}
\sim h$. To treat the gap breakdown process, we are actually looking 
for the minumum of $h$, which could be obtained by setting the 
derivative $h$ with respect to $l_{\rm acc}$ to zero. 
With some approximations, we finally get the 
gap parameters as $h=5.0\times 10^3({\rm cm})\zeta^{-3/7}P^{3/7}
B_{12}^{-4/7}\rho_6^{2/7}$, and $\Delta V=1.6\times 10^{12}
({\rm Volts})\zeta^{1/7}P^{-1/7}B_{12}^{-1/7}\rho_6^{4/7}$, 
which are analogous to RS75 (their eqs.[22],[23] except for the 
$\zeta$ correction), who treated the problem by simply adopting 
$h\sim l_{ph}$. By setting $\Delta V=\Phi_{\rm max}$(V), we get
the death lines of this model ($\zeta\sim 0.85$ has been adopted)
(cf. eqs.[6],[8] of CR93)
\begin{eqnarray}
\log \dot P  =(11/4)\log P-14.62 & [{\rm I}] \\
\log \dot P  =(9/4)\log P-16.58+\log\rho_6 & [{\rm I^{'}}] 
\end{eqnarray}
for the dipolar and the $\rho\sim R$, $B_s\sim B_d$ multipolar 
field configuration, respectively.

For the resonant ICS-induced gap, we have $E_c=2\gamma\hbar
(eB/mc)$ (Zhang et al. 1997; HM98), and
$\dot\gamma_{res}\sim 4.92\times 10^{11}B_{12}^2T_6\gamma_c
^{-1}$ (Dermer 1990), so that $l_e\sim 0.00276 \gamma_c^2 
B_{12}^{-1} T_6^{-1}$. 
Since $l_{\rm acc}\ll l_e\sim l_{ph}$ for the ICS case, we
solve the gap height by setting $h\sim l_{ph}\sim l_e$.
Also treating $T_6$ self-consistently through self-sustained 
polar cap heating, i.e., $T=(e\Delta V \dot N/\sigma\pi r_p^2)
^{1/4}$, where $\dot N=(\Omega B/ 4\pi e)
\pi r_p^2$ is the polar cap luminosity, and $r_p$ is the polar 
cap radius, we finally get the gap height $h=2.6\times 10^4
({\rm cm})P^{1/7}B_{s,12}^{-11/7}\rho_6^{4/7}\zeta^{-1/14}$, 
the gap potential $\Delta V=4.2\times 10^{13}({\rm Volts})
P^{-5/7}B_{s,12}^{-15/7}\rho_6^{8/7}\zeta^{6/7}$, and the
surface temperature $T_6=5.9 B_{12}^{-2/7}P^{-3/7}\rho_6^{2/7}
\zeta^{3/14}$ (Note that such a high polar cap temperature 
conflicts with the observations). The death lines of this model 
are then\footnote{The typical ICS photon energy adopted here is 
the one for resonant scattering. In the cases of high temperatures, 
the scatterings above the resonance with the photons at Planck's 
peak may become important (Zhang et al. 1997), and the death lines 
could be lower.}
\begin{eqnarray}
\log\dot P =(2/11)\log P-13.07 & [{\rm II}] \\
\log\dot P =(-2/11)\log P-14.50+(8/11)\log\rho_6 & [{\rm II}^{'}]
\end{eqnarray}
for dipolar and multipolar configurations, respectively.

\centerline{}
\centerline{\psfig{file=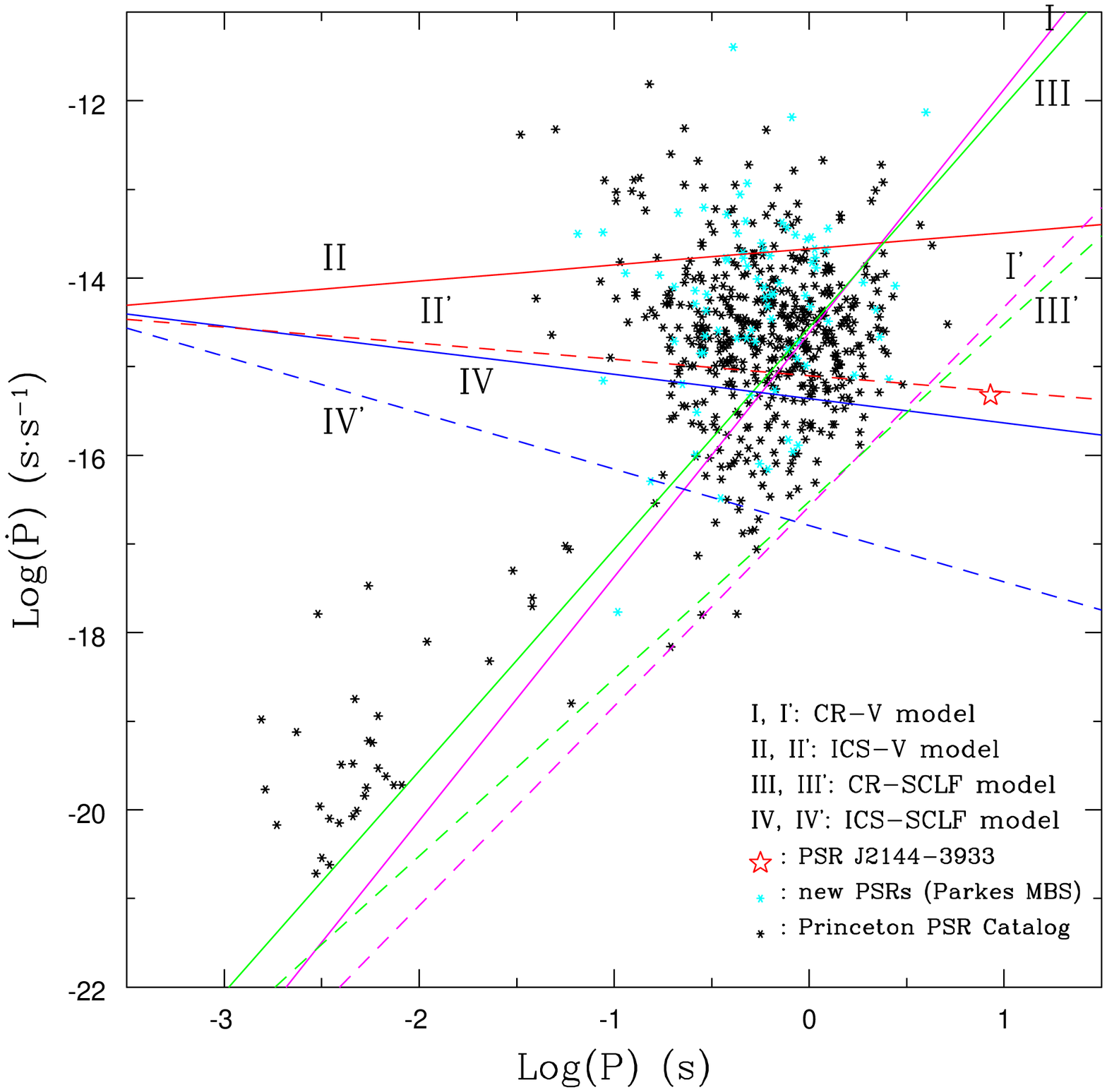,width=8.0cm}}
\figcaption{Death lines in various models. Solid lines are for dipole,
and dashed lines are for $\rho\sim 10^6$cm, $B_s\sim B_d$
multipole. Other captions are marked in the plot. Pulsar data follow
the latest results (Manchester et al. 1999).}
\centerline{}

\subsection{Space-charge-limited flow model (SCLF model)}

If the charged particles (electrons or ions) can actually be 
pulled out freely from the neutron star surface (which is favored 
by ion binding energy calculations), a space-charge-limited flow 
is a natural picture, with $E_\parallel=0$ at the surface. The 
$E_\parallel$ at higher altitudes then arises from the small 
imbalance of the local charge density $\eta$ from $\eta_{R}$ 
(eq.[\ref{etaR}]) due to the flow of the charged particles along
field lines.
The conservation of current requires $\eta\propto r^{-3}$,
thus the deviation $(\eta-\eta_R)$ arises from the extra 
$r$-dependence of $\eta_R$ (besides $B$ declination, which is 
$\propto r^{-3}$). In flat spacetime, this dependence is just the 
``flaring'' of the field lines, on which the AS79's pioneering SCLF 
model is based. Since such a deviation is so 
small, it takes a long length scale for a particle to be accelerated 
to pair producing energy via a gradual built-up of $E_\parallel$, so 
that the gap shape is usually narrow and long. The maximum
potential available in this model is much smaller than the one
available in V models (see footnote 9), so that
the death lines in this model are very high (see eq.[48], [50] of 
AS79). The ICS-induced version of such a SCLF model was presented by 
Luo (1996). HM98 explicitly studied both CR- and ICS- controlled
SCLF accelerators with the frame-dragging effect included, and
such a model is very good in interpreting high energy radiation
luminosities of the spin-powered pulsars (Zhang \& Harding 2000,
hereafter ZH00). The inclusion of the general relativistic 
frame-dragging effect in such SCLF models (MT92; MH97) is essential 
in two respects. First, besides the $B$ declination, $\eta_R$ in curved 
spacetime has an extra $(R/r)^3$ dependence (see second term 
in the bracket of eq.[\ref{etaR}]), while the current conservation 
requirement leads to $\eta \simeq -{{\bf \Omega_{*}\cdot B} \over 
2\pi c \alpha}(1-\kappa_g)$ near the surface. As a result, 
$E_\parallel$ is built up much faster. Secondly, the maximum
potential available is much larger than the flat spacetime value
(but smaller than that in V models), which is\footnote{For a full
expression of $\Phi_{\rm max}({\rm SCLF})$, see eq.[54] of HM98.
What we adopted here is the $\cos\chi$ term, which is much larger
than the $\sin\chi$ term (the maximum available for the AS79 model).}
\begin{equation}
\Phi_{\rm max}({\rm SCLF})\simeq \kappa_g{B_dR^3\Omega_{*}^2 \over 
2c^2}\simeq 1.0\times 10^{12}({\rm Volts})B_{d,12}P^{-2}R_6^3.
\label{Phimax2}
\end{equation}
As a result, the death lines are considerably lower than the AS79 model.

By introducing an upper $E_\parallel=0$ boundary at the pair 
formation front, HM98 have presented the explicit formalism and 
detailed numerical simulation of $E_\parallel$ within SCLF
accelerators. Unfortunately, simple analytic formulae applicable 
for all the cases are not available. However, we notice that near 
the death lines, the height of the accelerators ($h$) are all larger 
than the polar cap radius $r_{pc}=[\Omega_{*}R/cf(1)]^{1/2}R$ ($f(1)
\sim 1.4$ is the factor of curved spacetime), and  
that the $E_\parallel$ in the accelerators has achieved saturation
(see ZH00 for discussions of different regimes of acceleration
$E_\parallel$ and the criterion to separate the two regimes,
their eq.[53]). In such a case, we can adopt the following approximate 
acceleration picture: near the star surface, $E_\parallel$ grows
approximately linearly with respect to the height $z$ (thus is 
analogous to V model in this regime) with the form $E_\parallel(1)
\simeq [3\kappa_g/(1-\epsilon)](\Omega_{*}B/c)z \simeq 157B_{12}
P^{-1}z$ (eq.A[3] of HM98), where $\epsilon=r_g/R\sim 0.4$, and 
saturates above $z\sim r_{pc}/3$ at a value of $E_\parallel(2)\simeq 
(3\kappa_g/2)({\bf \Omega_{*}\cdot B}/ c)r_{pc}(r_{pc}/R)(1-\xi^2)
\simeq 3.5\times 10^3 B_{12}P^{-2}R_6^2$ (eq.[A5] of HM98), where
$f(1)=1.4$, $\xi=0.7$ have been adopted. To study the death lines, 
we assume that the accelerators are located at the surface, though 
HM98 have argued that accelerators could be $(0.5-1)R$ above the
surface in young pulsars due to anisotropy of the upward versus 
downward ICS. In old pulsars, the returning positron fraction is
smaller, and the lower pair formation front may not exist, so 
that both the CR- and the ICS- induced SCLF accelerators could 
be formed at the surface.

For the CR-induced model, we again have $l_e\ll l_{\rm acc}\sim 
l_{ph}$. Since $h>r_{pc}$ near the death lines (ZH00), we can adopt
$\gamma_c\sim (e/mc^2)E_\parallel(2)l_{\rm acc}$. Following the 
same procedure as for the V model, but using $\Delta V
=E_\parallel(2)h$ (recall the quadratic form of V model and
note the difference), we get $h=3.3\times 10^5 ({\rm cm})
P^{3/2}B_{12}^{-1}\rho_6^{1/2}R_6^{-3/2}$ and $\Delta V=3.5\times
10^{11}({\rm Volts})P^{-1/2}\rho_6^{1/2}R_6^{1/2}$. Equating 
$\Delta V$ with $\Phi_{\rm max}$(SCLF) (eq.[\ref{Phimax2}]), we
get the death lines 
\begin{eqnarray}
\log\dot P =(5/2)\log P-14.56 & [{\rm III}]\\
\log\dot P =2\log P-16.52+\log\rho_6 & [{\rm III^{'}}]
\end{eqnarray}
for dipolar and multipolar field configurations, respectively.

For the resonant ICS-induced SCLF model, we again have $l_{\rm acc}
\ll l_e\sim l_{ph}$ and $h>r_{pc}$. This brings an important 
difference with the CR-induced case, that is, one should 
adopt the linear $E_\parallel$ form, e.g., $\gamma_c=(e/mc^2)
\int_0^{l_{\rm acc}}E_\parallel(1)dz$ to describe the acceleration 
phase, but adopt the saturated $E_\parallel$ form to describe 
the final potential, i.e., $\Delta V=E_\parallel(2)h$. Since
the SCLF model has a much lower charge deficit than the V model, 
the number of reversed positrons required to screen the field is 
only a factor of $f\simeq \left| (\partial E_\parallel/\partial r)
/(8\pi \rho_{_{GJ}}) \right|$ of the Goldreich-Julian density 
(AS79; ZH00), so that one gets less polar cap heating, i.e., 
$T=(e\Delta V\dot N f/\sigma\pi r_p^2)^{1/4}$, in this model. 
For the saturated accelerators, this factor is roughly
$f\simeq 5.7\times 10^{-5}P^{-1}$ near the surface (eq.[71]
of ZH00). A self consistent treatment of $T$ finally leads
to $h=9.7\times 10^4({\rm cm})P^{4/13}B_{12}^{-22/13}
\rho_6^{8/13}R_6^{-2/13}$, $\Delta V=1.0\times 10^{11}({\rm Volts})
P^{-22/13}B_{12}^{-9/13}\rho_6^{8/13}R_6^{24/13}$, and $T_6=0.11 
P^{-12/13}B_{12}^{1/13}\rho_6^{2/13}R_6^{6/13}$, so that the death 
lines are
\begin{eqnarray}
\log\dot P =-(3/11)\log P-15.36 & [{\rm IV}]\\
\log\dot P =-(7/11)\log P-16.79+(8/11)\log\rho_6. & [{\rm IV^{'}}]
\end{eqnarray}
In this model, the polar cap temperature is sustained by the thermal
energy released from the crust deposited by the reverse flow of
positrons and high energy photons. The primary beam fluctuations
(typical timescale $\sim h/c \sim 3\times 10^{-6}$s)
may result in discontinuous illumination of the polar cap. However,
because of the ``inertia'' of photon diffusion from the relatively
deep layers to the photosphere (e.g. Eichler \& Cheng 1989), this 
process is unlikely to affect the average polar cap temperature. For 
example, it takes $\sim 6\times 10^{-3} T_5^{-1.5}$s
for the photons
to diffuse upthrough the surface from a depth of 100${\rm g~cm}^{-2}$.

\section{Conclusion and discussions}
The death lines of different models are plotted in Fig.1. A
remarkable fact is, though PSR J2144-3933 is beyond the death valleys 
of both the CR- and ICS- induced V models and CR-induced SCLF model, 
it is well above the {\em star-centered dipolar} death line of the 
ICS-induced SCLF model. Thus one does
not need to introduce a special neutron star equation-of-state or
anomalous field configurations at all to maintain strong pair formation 
in this pulsar. The ICS-SCLF death 
lines of Fig.1 imply a very large phase space for radio emission in 
the long period regime, thought previously to be radio forbidden. In fact, 
Young et al. (1999) argued that the population of pulsars with
parameters 
silimar to PSR J2144-3933 is very large, since the detectability of 
such pulsars is very small due to their small polar caps. We expect more 
pulsars to be detected in this region. A general trend in Fig.1 is that 
death lines in the ICS-induced models have much flatter slopes than those 
in the CR-induced model. This arises from the very different $P$-, $B_s$- 
dependences of both $l_{ph}$ and $l_e$ for the resonant ICS processes. 
The saturation of $E_\parallel$ in long period pulsars is critical in
lowering the death lines in the SCLF models relative to those of the
V models\footnote{V models also have $E_\parallel$ saturation, but only 
near the death lines when $h\sim r_{pc}$ (RS75). SCLF models, however, 
can have $E_\parallel$ saturation much farther away from the death lines
and thus allows a greater distance for pair production before 
$\Phi_{\rm max}$ is reached.}. The difference is more prominant in ICS-
induced models. For the ICS-SCLF model, our death lines are lower than
the ones derived in Arons (2000). The difference might be due to the
different polar cap temperature treatments.

We thank the referee for good comments and suggestions and Demosthenes 
Kazanas, G.J.Qiao, R.X.Xu, and Z.Zheng for interesting discussions or
helpful comments.

\end{document}